\documentclass[journal]{IEEEtran}
\usepackage{color}
\usepackage{multicol}
\usepackage{graphicx}
\usepackage{multirow}
\usepackage{booktabs}
\usepackage[table]{xcolor}
\usepackage{algorithm}
\usepackage{algorithmic}
\usepackage{amsmath,amssymb,amsthm}
\usepackage{epic,multibox,fancybox}
\usepackage{float}

\begin{document}

\title{AM-DisCNT: \underline{A}ngular \underline{M}ulti-hop \underline{DIS}tance based \underline{C}ircular \underline{N}etwork \underline{T}ransmission Protocol for WSNs}

\author{A. Rao$^{1}$, M. Akbar$^{1}$, N. Javaid$^{1,2}$, S. N. Mohammad$^{1}$, S. Sarfraz$^{1}$\\\vspace{0.4cm}
$^{1}$Dept of Electrical Engineering, COMSATS Institute of IT, Islamabad, Pakistan.\\
$^{2}$CAST, COMSATS Institute of IT, Islamabad, Pakistan.}

\maketitle
\begin{abstract}
 The nodes in wireless sensor networks (WSNs) contain limited energy resources, which are needed to transmit data to base station (BS). Routing protocols are designed to reduce the energy consumption. Clustering algorithms are best in this aspect. Such clustering algorithms increase the stability and lifetime of the network. However, every routing protocol is not suitable for heterogeneous environments. AM-DisCNT is proposed and evaluated as a new energy efficient protocol for wireless sensor networks. AM-DisCNT uses circular deployment for even consumption of energy in entire wireless sensor network. Cluster-head selection is on the basis of energy. Highest energy node becomes CH for that round. Energy is again compared in the next round to check the highest energy node of that round. The simulation results show that AM-DisCNT performs better than the existing heterogeneous protocols on the basis of network lifetime, throughput and stability of the system.
\end{abstract}

\begin{IEEEkeywords}
Wireless sensor networks, static clustering, specifically randomly deployed, uniformly randomly deployed, clustering algorithm, routing protocols.
\end{IEEEkeywords}
\IEEEpeerreviewmaketitle
\section{Introduction}

WSNs are highly distributed networks of small, compact and lightweight sensors called Nodes. These nodes are deployed numerously to monitor the environment conditions such as temperature, light, sound, fire etc. After processing the gathered data, node sends its data to BS. Applications of WSNs include constant monitoring and detection of specific events such as military and battlefield surveillance, fire forecast, flood detection and patient monitoring etc.

 Some design issues and challenges should be kept in consideration in WSNs. In most of WSN protocols nodes are deployed randomly. Random deployment in larger areas leaves some regions unmonitored. Which is a fundamental issue in random deployment. It reflects how well an environment is being monitored.
 Energy is a major constraint in WSNs. Each node has an energy source; usually a battery. In most cases, it is very difficult and sometimes impossible to re-energize the batteries of a node for further use. Therefore, it is very challenging to design protocols that minimize energy consumption.  The maintenance of an autonomous system i.e. a network consisting of a number of different nodes sharing similar data, is almost impossible. A trade off always exists between the different design constraints. A major trade-off exists between network lifetime and packets to the main monitoring entity. BS placement should be such that it can get packets from every part of network conveniently. In the proposed protocol, we have tried to optimize these parameters. Due to the energy constrained, such routing protocols are needed to utilize minimum energy during communication. DC nodes send their data directly to BS therefore, lifetime of the nodes nearer to the BS increases. DC (Direct communication) is not a good technique for farther nodes, as they die quickly. For the farther nodes clustering is required to preserve energy. So, AM-DisCNT is designed to take the advantage of both DC and cluster formation. In this protocol, inner circle nodes communicate directly with the BS, whereas, outer circle nodes form clusters in their defined areas. CHs gather data from the nodes associated with them and after aggregation, send data to BS using multi-hop technique.

\section{Related Work}
 In this section a brief over view of related research work is given.
 There are two types of clustering schemes: homogeneous clustering scheme and heterogeneous clustering scheme.
 Under homogeneous clustering scheme LEACH~\cite{1} i.e. Low energy adaptive clustering hierarchy was proposed by Heinzelman. It is a hierarchical clustering algorithm that randomly selects nodes as CHs. The protocol is based on the initial energy of the system.  In LEACH every node has a specified probability to become a cluster head. LEACH uses two modes for communication i.e. between nodes and CHs and between CHs and BS. Node takes the decision to become a CH by generating a random number between 0 and 1.

  LEACH-Centralized i.e. LEACH-C~\cite{2},  a protocol proposed by Heinzelman and Balakrishnan, uses centralized clustering algorithm. In LEACH-C all the information about energy and location of nodes is send to base-station. The CH selection is random in LEACH-C.

 Multihop-LEACH~\cite{3}, a protocol proposed by N. Israr and I. Awan, contain two operations i.e. Multihop- inter cluster and Multihop-intra cluster. The data send by nodes is then received at the BS through a chain of CHs. The farther nodes send a request to their nearer nodes to become their temporary CH.

  TEEN (Threshold sensitive energy efficient sensor network protocol)~\cite{4} was proposed by Manjeshwae. TEEN defines two thresholds: hard and soft threshold.  Data is not transmitted until the threshold is reached. That’s why TEEN is not a good option where data is needed periodically.

 APTEEN (Adaptive Threshold sensitive Energy Efficient sensor Network protocol)~\cite{5} provides data periodically and also provides information on time-critical events. The architecture resembles to that of TEEN.

  PEGASIS (Power-Efficient GAthering in Sensor Information Systems)~\cite{6} is based on chain formation instead of cluster formation. PEGASIS uses it’s chain head to transmit the data of whole chain to BS.

 In heterogeneous clustering scheme, Md. Solaiman proposed ALEACH (Advanced LEACH)~\cite{7} which elects the cluster head depending on both the current state and random probability.

  The performance parameters such as 'Network lifetime', 'Stability period' and 'Throughput of the system' are calculated by keeping an eye on the work of authors in ~\cite{8}, ~\cite{9}, ~\cite{10} and ~\cite{11}. Whereas, the number of alive and dead nodes are calculated by keeping in mind the work of authors in ~\cite{12}.

 SEP (Stable Election Protocol)~\cite{13}  by G. Smaragdakis have two level heterogeneous hierarchical network, In which every node independently elects itself as a cluster head based on initial energy compared to other nodes’ energy.

  DEEC~\cite{14} (Distributed energy efficient Clustering) proposed by Li Qing shows the algorithm in which cluster head selection is based on the probability of ratio of residual energy and average energy of the network. DEEC out performs homogeneous schemes and SEP as well.

 SDEEC ~\cite{15} (Stochastic Distributed Energy-Efficient Clustering) by B. Elbhiri introduced a balanced method for CH election. This method is more efficient than previous techniques as it uses stochastic scheme detection. SDEEC out performs SEP and DEEC in terms of network lifetime.
  TDEEC (threshold distributed energy efficient clustering) increases the stability of heterogeneous network.


\section{Radio Energy Dissipation Model}
AM-DisCNT considers a simple radio model adopted from ~\cite{16}. Energy dissipates to turn on the transmitter or receiver. In first order radio model $E_{elec} = 50 nJ/bit$ and $E_{amp} = 100 pJ/bit/m^2$. $E_{amp}$ is used to achieve an acceptable SNR ratio.
We also consider a path loss of $(distance)^2$ in transmitting a ‘k’ bit packet. Equation for transmission of ‘k’ bit packet through distance ‘d’, provided $d<d_{0}$ is:

 \begin{equation}
 E_{TX} = E_{elec}k+E_{amp}kd^2
 \end{equation}
 Equation for transmission of $‘k’$ bit packets through distance $‘d’$, provided $d \geqslant d_{0}$ is:
 \begin{equation}
 E_{TX} = E_{elec}k+E_{amp}kd^4
 \end{equation}
  For the reception of ‘k’ bit packet, equation is;

 \begin{equation}
 E_{RX} = E_{elec}k
 \end{equation}

\begin{figure}[ht]
\begin{center}
\includegraphics[height=3cm, width=5.5cm]{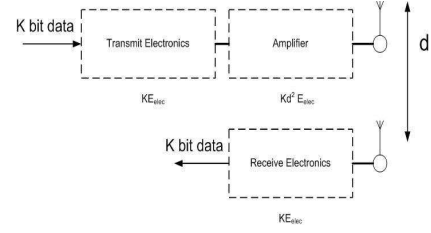}
\caption{Radio Dissipation Model}
\end{center}
\end{figure}
We assume symmetry in our radio channel. We also assume that nodes always have data to send to the BS. Therefore, from above equations we can conclude that along with distance between the nodes and BS, the protocol should also minimize the number of transmit and receive operations in the network to preserve energy.
\section{AM-DisCNT}
 This section explains the detailed functioning of the protocol. AM-DisCNT uses DC and static clustering to cope with the design constraints. The cluster head selection is on the basis of energy. AM-DisCNT improves the stability period and network lifetime by conserving the energy consumption.
 The following subsection describes the distribution of network area.
\subsection{AM-DisCNT field distribution}
 AM-DisCNT divides the whole network into two concentric circles. Inner circle with radius R1 and outer circle with radius R2. BS is placed at the centre of circle. Circular region is considered, to get maximum output from every region of network. Unlike rectangular networks, the corner nodes will not consume extra energy during communication. Inner circle nodes are supposed to send their status directly to the BS due to minimum distance. The communication of outer circle nodes is done through cluster formation.
\subsection{AM-DisCNT Architecture}
 In this section the schematic diagram of AM-DisCNT is summarized. ‘N’ number of nodes are deployed randomly in two circular regions i.e. inner circle and outer circle. Nodes are taken as micro-mobile or stationary i.e. once deployed, they cannot move or change their position. Inner circle nodes are directly associated with BS. Whereas, the outer circle nodes are further divided into eight regions. This physical division is done for the purpose of clustering. Equal numbers of nodes are deployed in each region. In the outer eight regions, nodes are uniformly randomly deployed to provide full coverage area.
 Nodes are often located far from the base station and they always have data to transmit to the BS. Such nodes can be used to monitor the environment or they can be useful in military operations. Keeping it in mind we suppose the BS as static sink and is placed in the center of the two concentric circles.
\begin{figure*}[ht]
\begin{center}
\includegraphics[height=9cm, width=8cm]{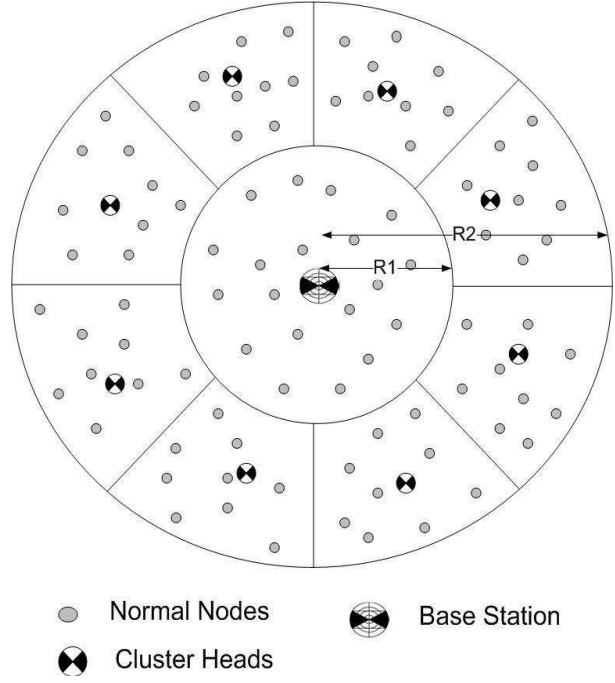}
\caption{AM-DisCNT Schematics}
\end{center}
\end{figure*}

\begin{figure}[ht]
\begin{center}
\includegraphics[height=5.5cm, width=5.5cm]{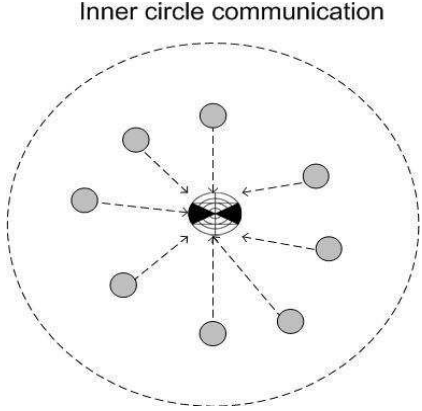}
\caption{Type1 communication }
\end{center}
\end{figure}

\begin{figure}[ht]
\begin{center}
\includegraphics[height=5.5cm, width=5cm]{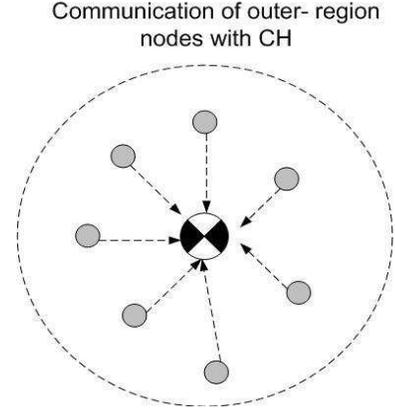}
\caption{Type2 communication }
\end{center}
\end{figure}

\begin{figure}[ht]
\begin{center}
\includegraphics[height=7cm, width=8.5cm]{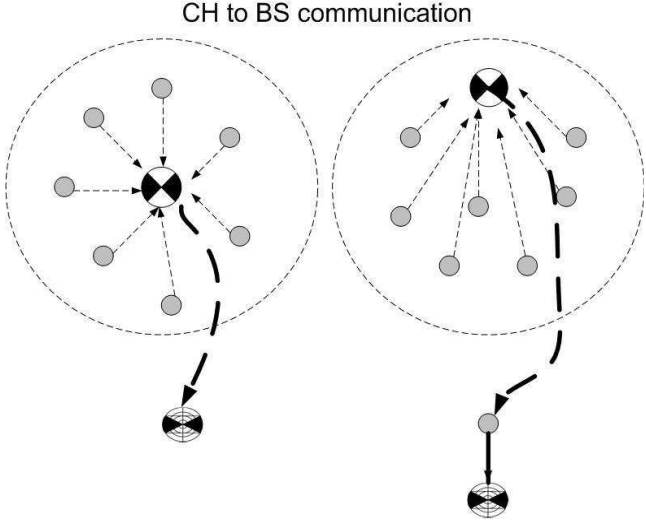}
\caption{Type3 communication }
\end{center}
\end{figure}

\begin{figure}[ht]
\begin{center}
\includegraphics[height=5cm, width=8cm]{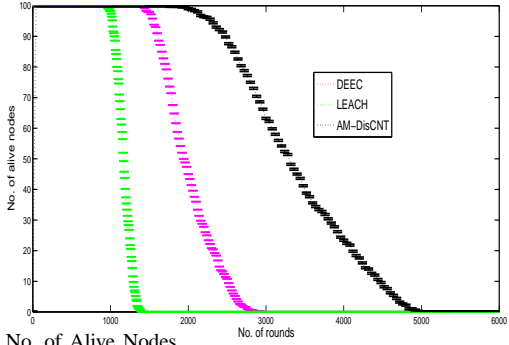}
\vspace{-0.7cm}
\caption{No. of Alive Nodes}
\end{center}
\end{figure}

\begin{figure}[ht]
\begin{center}
\includegraphics[height=5cm, width=8cm]{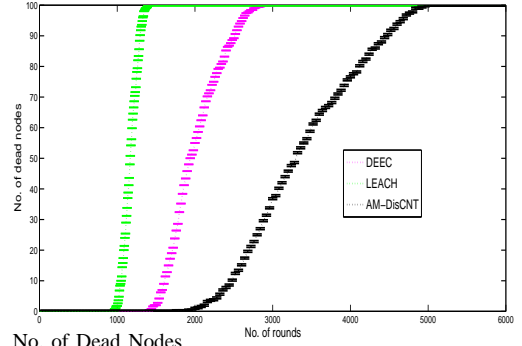}
\vspace{-0.7cm}
\caption{No. of Dead Nodes}
\end{center}
\end{figure}

\begin{figure}[ht]
\begin{center}
\includegraphics[height=5cm, width=8cm]{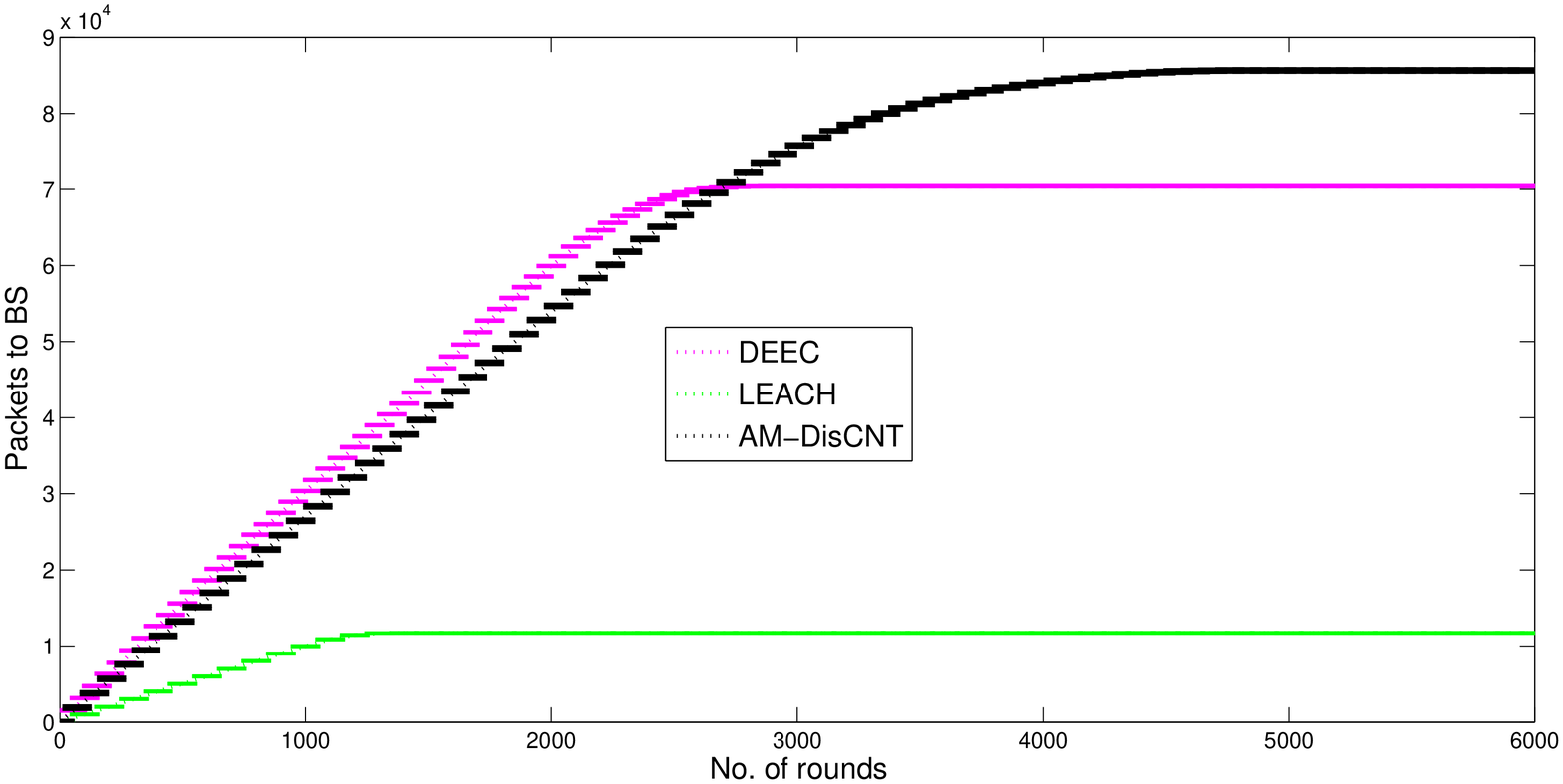}
\vspace{-0.7cm}
\caption{No. of packets sent to BS}
\end{center}
\end{figure}

\begin{figure}[ht]
\begin{center}
\includegraphics[height=5cm, width=8cm]{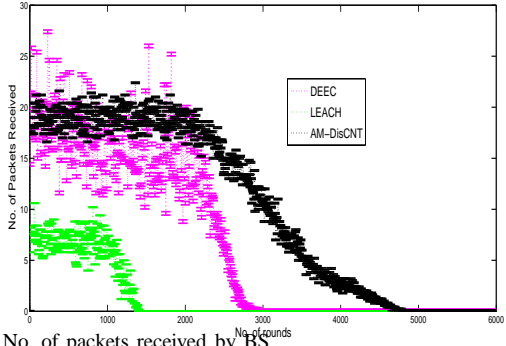}
\vspace{-0.7cm}
\caption{No. of packets received by BS}
\end{center}
\end{figure}

\begin{figure}[ht]
\begin{center}
\includegraphics[height=5cm, width=8cm]{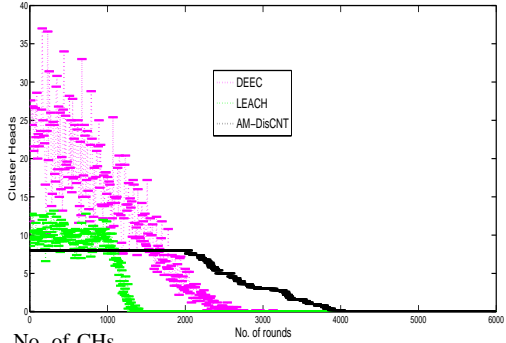}
\vspace{-0.7cm}
\caption{No. of CHs}
\end{center}
\end{figure}

\begin{figure}[ht]
\begin{center}
\includegraphics[height=5cm, width=8cm]{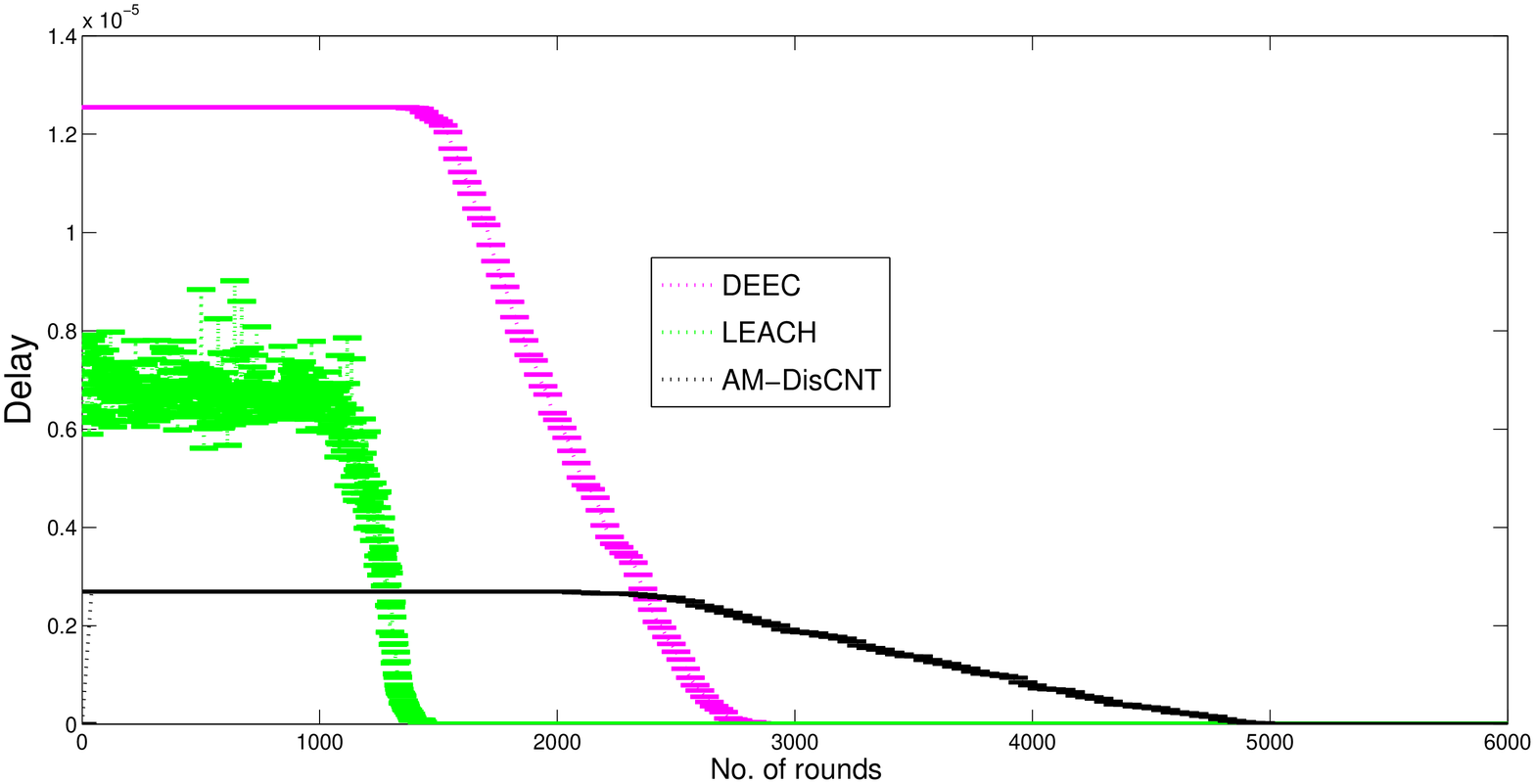}
\vspace{-0.7cm}
\caption{Delay}
\end{center}
\end{figure}

\subsection{AM-DisCNT Region formation}
 The outer circle of AM-DisCNT is divided into eight equal regions. Thus, the protocol consists of total nine regions i.e. inner circular region (T1) and eight outer regions (from R1 to R8). The division of regions is such that it decreases the distance of nodes from CH and from CH to BS. Circular region T1 of the network is formed to separate the nearer nodes from the farther nodes. 'N' number of nodes are uniformly randomly deployed in the network.
 In the region $T1$, nodes $n_{T1}$ are randomly circularly deployed. The x and y coordinate of nodes $n_{T1}$ can be calculated by using the equations as follows:

 \begin{equation}
 X_{cood} = R1\cos(\theta)
 \end{equation}

 \begin{equation}
 Y_{cood} = R1\sin(\theta)
 \end{equation}
 Where, $0\leq \theta\leq2\pi$ and $0<R1\leq S$. S can be any positive integer.

 In order to deploy the wireless sensor nodes, we assume the ability to detect the “empty areas” and then deploy nodes. In this case, first of all the nodes $n_{T1}$ are deployed in the region T1 then the nodes $n_{R1}$ are deployed in the region R1, nodes $n_{R2}$ are deployed in the region R2 and so on. We also assume that the communication range of all the wireless sensor nodes are within their own defined regions except CHs. The nodes are bound to communicate within their own specified regions.
 In the regions, R1 to R8, the x and y coordinates of nodes $n_{R1}$ to $n_{R8}$ are given by the following equations:
 \begin{figure}[ht]
\begin{center}
\includegraphics[height=5cm, width=8cm]{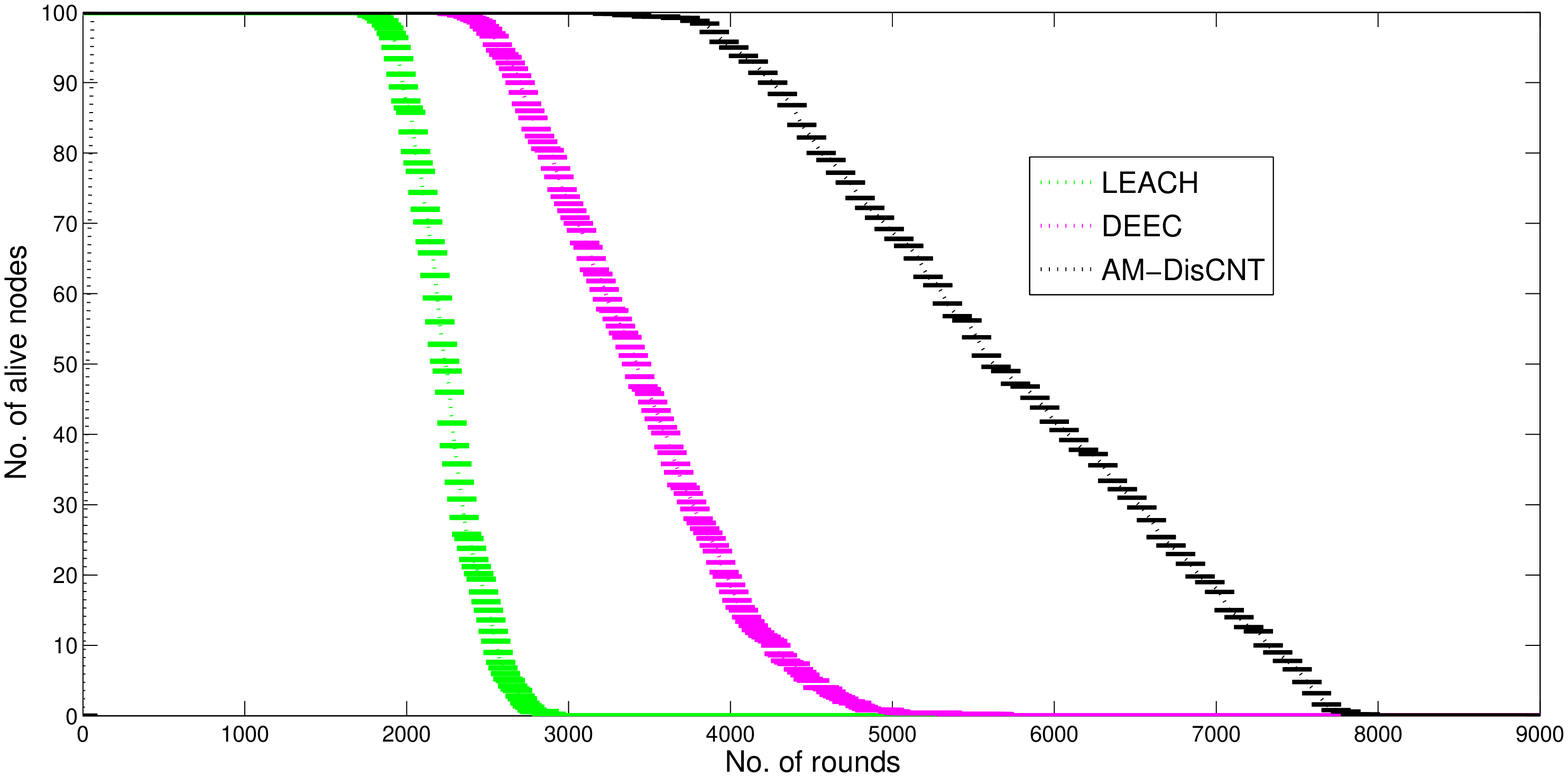}
\vspace{-0.7cm}
\caption{Alive nodes for parameters in Table 2}
\end{center}
\end{figure}

\begin{figure}[ht]
\begin{center}
\includegraphics[height=5cm, width=8cm]{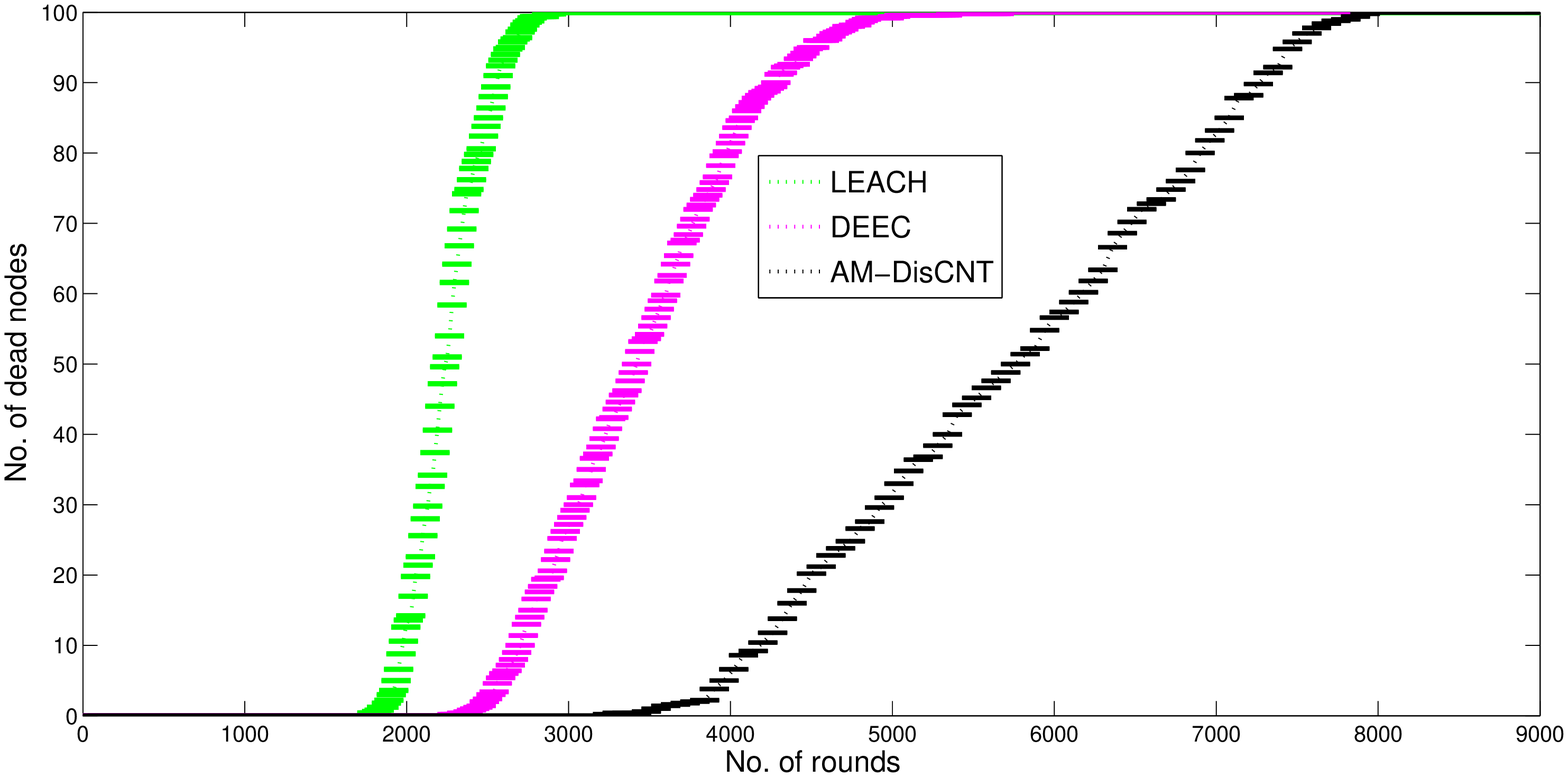}
\vspace{-0.7cm}
\caption{Dead nodes for parameters in Table 2}
\end{center}
\end{figure}

\begin{figure}[ht]
\begin{center}
\includegraphics[height=5cm, width=8cm]{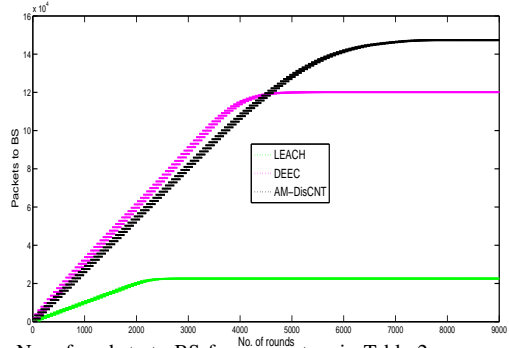}
\vspace{-0.7cm}
\caption{No. of packets to BS for parameters in Table 2}
\end{center}
\end{figure}

\begin{figure}[ht]
\begin{center}
\includegraphics[height=5cm, width=8cm]{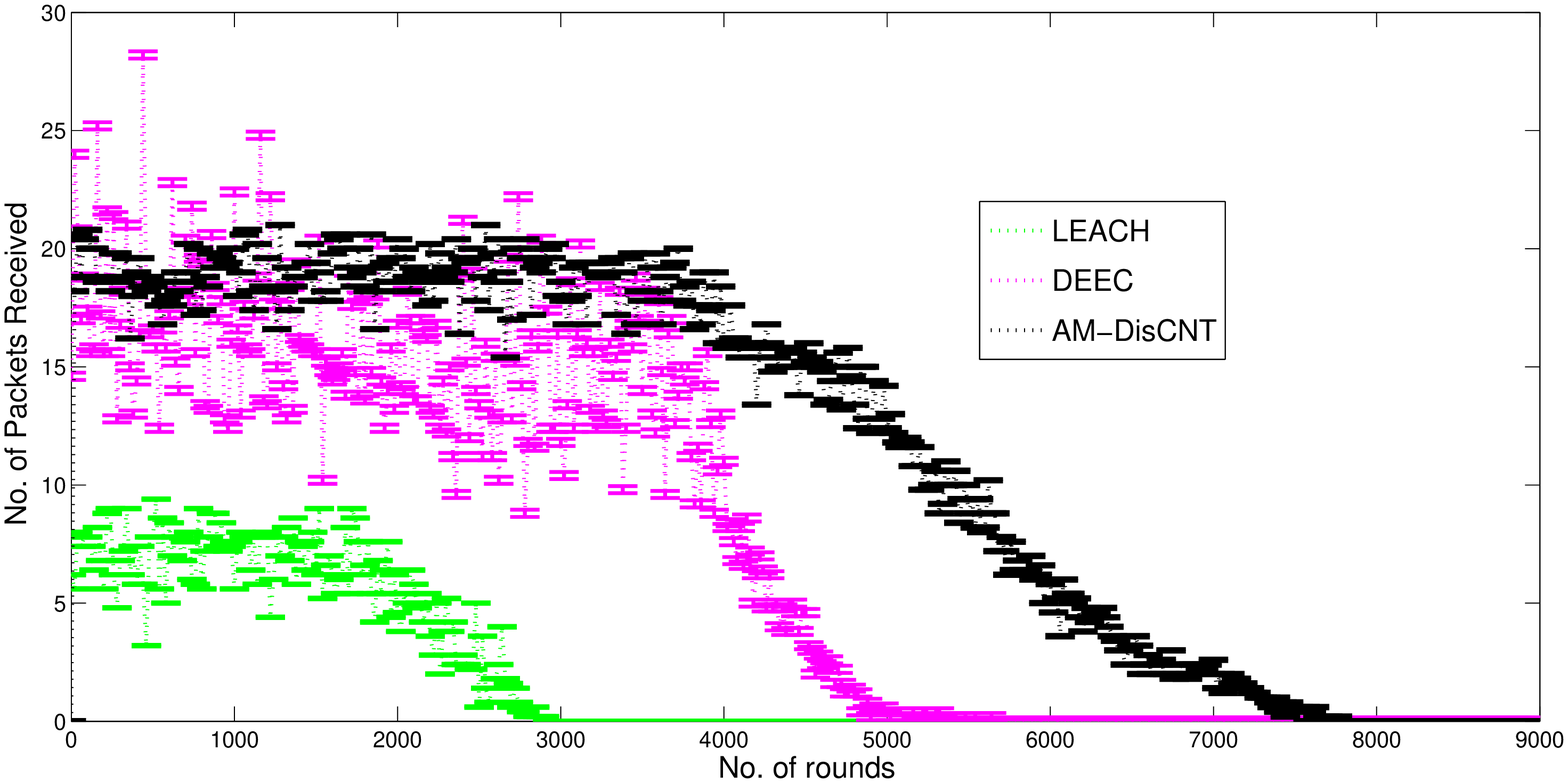}
\vspace{-0.7cm}
\caption{No. of packets received by BS for parameters in Table 2}
\end{center}
\end{figure}

\begin{figure}[ht]
\begin{center}
\includegraphics[height=5cm, width=8cm]{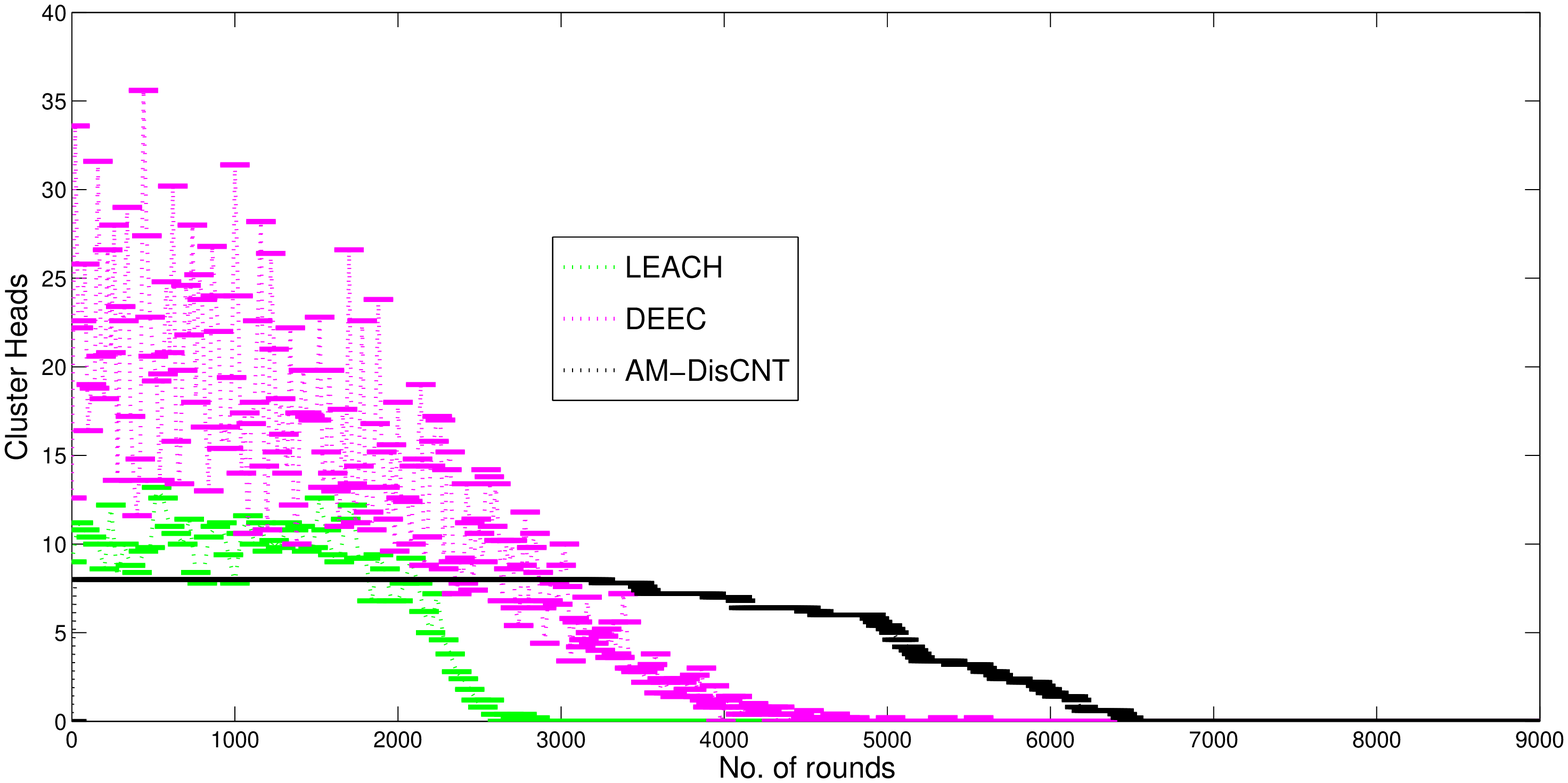}
\vspace{-0.7cm}
\caption{Cluster Heads with changed parameters refer to Table 2}
\end{center}
\end{figure}

\begin{figure}[ht]
\begin{center}
\includegraphics[height=5cm, width=8cm]{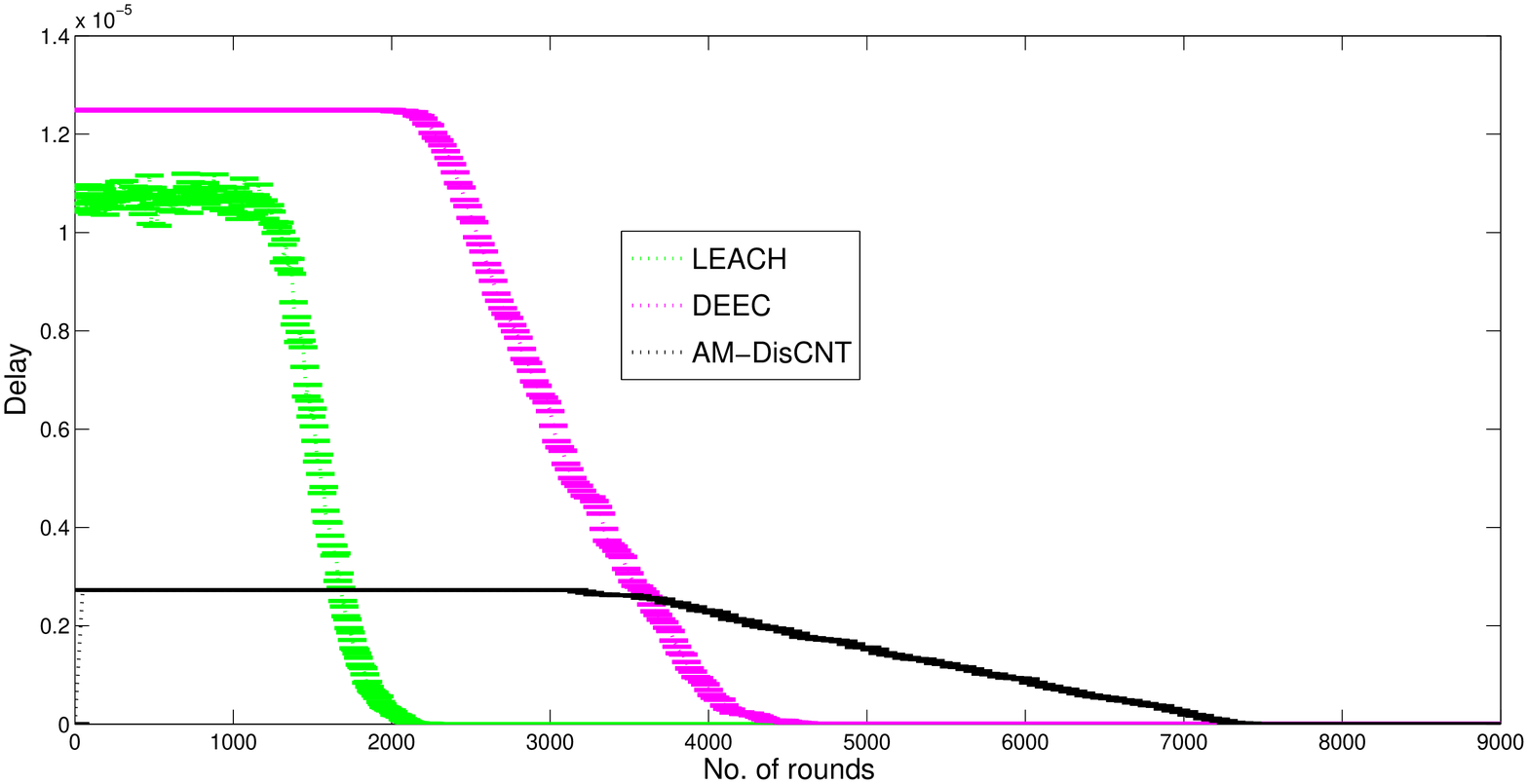}
\vspace{-0.7cm}
\caption{Delay with changed parameters refer to Table 2}
\end{center}
\end{figure}

 \begin{equation}
 X_{cood} = R_{0}\cos(\theta)
 \end{equation}

  \begin{equation}
 Y_{cood} = R_{0}\sin(\theta)
 \end{equation}

 Here, theta is a variant of $n(\pi/4)$, given $n=0$ to 8. $R_{0}= R2-R1$.
 To define a single region in outer circle, two limits of theta are used.

\subsection{Cluster head selection}
 CH selection is on the basis of maximum energy. Highest energy node in each of the defined region becomes the CH for that round. CH collects the data of its own region and after aggregation sends it to BS. The transmission from CH to BS is done taking the advantage of multi-hop. After the first round, energy of each node is checked again and highest energy node is selected as CH and so on. Such a type of clustering ensures full coverage area and optimized energy utilization.

\subsection{Heterogeneity of network}
 In this section we will discuss the heterogeneity of our network. In two level heterogeneous networks, two types of sensor nodes are used i.e. normal and advanced nodes. Advanced nodes own $\alpha$ times more energy than normal nodes. $E_{0}$ is the initial energy of normal nodes and $(1-m)N$ is the total number of advanced nodes. mN number of advanced nodes are equipped with total energy of $E_{0}(1+\alpha)$. Thus, the total initial energy of two level heterogeneous network is given by the following equation;
\begin{equation}
E_{total} = E_{0}(1-m)N + mNE_{0}(1+\alpha)
\end{equation}
\begin{equation}
E_{total} = NE_{0}(1+m\alpha)
\end{equation}
 Total energy of three level heterogeneous networks is given by;
 \begin{equation}
E_{total} = NE_{0}(1+m(\alpha+m_{0}\beta))
\end{equation}

 Where, the fraction of super nodes is denoted by $m_{0}$, m is the fraction of advanced nodes and the super nodes have $\beta$ times more energy than normal nodes.

 AM-DisCNT considers a wireless multi-level heterogeneous network. Energy is randomly distributed among all the sensor nodes of the network. The energy of sensor nodes is given by the equation;
\begin{equation}
E_{node} = E_{0}(1+ t\alpha)
\end{equation}
 Where, $E_{0}$ is the initial energy of nodes and the nodes may have $\alpha$ times more energy than the initial energy $E_{0}$. Thus, the total energy of multi-level heterogeneous wireless network is given by;
\begin{equation}
E_{total} = \sum\limits_{t=1}^N E_{0} (1+ t\alpha)
\end{equation}

\section{Performance Evaluation}
  Our protocol is compared with LEACH and DEEC. In multi-level heterogeneous  networks, the extended protocols of LEACH and DEEC are used. The following scenarios are considered for performance measures. Where, $\alpha=1$ , $R1=20$ and $R2= 35$. The simulation results show that the performance of LEACH under multi- level heterogeneous network is increased. DEEC is more stable than LEACH. It is obvious from results that the stable period of AM-DisCNT is prolonged compared to that of LEACH and DEEC. It is because the advanced nodes die more slowly.

 Following are the parameters that are considered in this protocol for evaluation:
 \begin{itemize}
  \item \emph{Network lifetime}: It is the time period from the start of first round to the death of last node.
  \item \emph{Stability period}: It is the time interval or number of rounds covered till death of first sensor node.
  \item \emph{Dead nodes}: The number of nodes that can't send their data to BS: contribution in network is nullified.
  \item \emph{Alive nodes}: Nodes that have some residual energy and they can send their status to CH or BS are termed as alive nodes.
  \item \emph{Network throughput}: Throughput of the network is the number of data packets that are send to base-station.
  \item \emph{Confidence interval}: It is used to indicate the reliability and authentication of estimated results.
  \item \emph{Packets received}: Packets received refers to the number of packets received at the BS.
  \item\emph{Delay}: The time consumed by the packet to reach the specified destination is termed as delay.
 \end{itemize}
 Obviously, greater the stability, better will be the reliability of wireless sensor network but there is always a tradeoff between network lifetime and reliability.Stable region determines the reliability of network. Even after the death of few nodes, base-station receives sensed data from network but this information is not trustable. Keeping it in mind, this protocol is designed for optimal number of CHs to get maximum coverage area and reliable information.

 MATLAB is used as a simulator to analyze performance of cluster base routing protocol . A circular network is considered having area equal to the area of outer circle. $N=100$ nodes are randomly deployed in their specified areas. BS is placed at the center of two concentric circles. We took the results by running each simulation five times and plotted their average and confidence intervals. All the parameters taken for these simulations are given in the TABLE I.
 \begin{table}[htbp]
  \centering
  \caption{Network Parameters}
  \begin{tabular}[height=9cm,width=9cm]{lll}
    \toprule
    Parameter & Value  &  \\
    \midrule
   Area of network          & $\pi r^2$ \\
   N            & 100  \\
  $E_0$          & 0.5 j \\
     $ETX$        & 50/nj/bit \\
     $ERX$       & 50/nj/bit \\
    $E_fs$       & $10/pj/bit/m^2$ \\
    $E_mp$       & $0.0013/pj/bit/m^4$\\
     $E_DA$       & $5/nj/bit$ \\
    $E_{elec}$     & $50nj/bit$ \\
     \bottomrule
 \end{tabular}%
 \label{tab:addlabe2}%
 \end{table}%

\subsubsection{Stability period}
The performance of AM-DisCNT is evaluated on the basis of stability period, compared with those of LEACH and DEEC. AM-DisCNT carry out more rounds from the start of network till the end of first node, hence greater  stability. Greater stability of AM-DisCNT is due to the even distribution of nodes in defined regions. Regarding stability, AM-DisCNT out performs both LEACH and DEEC.

\subsubsection{Network lifetime}
 Fig. 6 and Fig. 7 show that our proposed protocol out performs other protocols in terms of network life time as well.  The CHs in the outer region die first because of the large communication distance from BS, which is placed at the center of concentric circles. The nodes that are deployed in the inner circle are communicating directly with the BS, which will increase the number of rounds till their death. AM-DisCNT also takes the advantage of multi-hop concept. The CHs of outer region send their aggregated data to inner circle nodes that are at the minimum distance from the BS. This multi hopping takes place on the basis of distance. Hence, the network lifetime is increased.

\subsubsection{No. of CHs}
 Fig. 10 shows that instead of having variable CHs, we rather have fixed CHs per round i.e. one CH per region in the outer circle. This optimal selection of CHs per round ensures the data sent from every part of network to BS. These optimal numbers of CHs collect data from their associated nodes within the defined region and send it to BS.

\subsubsection{Confidence interval}
The confidence interval is the statistical estimate that is used to calculate the range of values that is likely to contain the data of interest. This interval is calculated using the formulae;
      \begin{equation}
  \bar{x}-\bar{z}(\alpha/2)\surd(\sigma/n) < CI <  \bar{x}+\bar{z}(\alpha/2)\surd(\sigma/n)
      \end{equation}

  The left side of equation gives the lower bound and right side is used to calculate the upper bound. $\sigma$ in the above equation is given by;
      \begin{equation}
  \sigma=\surd(1/N(\sum\limits_{i=1}^N (X_i-\bar{x})^2)
      \end{equation}
  Here, N is the number of alive nodes. $\bar{z}$ is the averaged value. $X_i$ are the samples of nodes(dead, alive etc).

 Fig. 8 depicts the successful packets send to the BS. The throughput of LEACH and DEEC is less than AM-DisCNT. The reason behind this is; the fixed number of CHs per region. Single CH per region implies the collection of data from all the associated nodes and transmitting it to BS after aggregation. This technique reasonably increases the stability. As the stability increases, the throughput of network also increases. The variable number of CHs in LEACH and DEEC does not insure the coverage of whole network and reduce the efficiency of network in terms of stability and lifetime. Inner circle nodes also play a significant role in increasing the throughput of the system.

 \subsubsection{Packets Received}
 Fig. 9 depicts the packets successfully received at BS. The received packets of LEACH and DEEC fluctuate more than AM-DisCNT. Packet drop rate of AM-DisCNT is less than its counterparts, this is because of the optimal communication distance.

 Fig. 11 shows the comparison of AM-DisCNT, LEACH and DEEC on basis of delay. Delay occurs when a packet is send to BS. From Fig. 11 we can interpret that the delay caused by AM-DisCNT is least as compared to LEACH and DEEC. This is due to the optimal clustering of AM-DisCNT.
\subsubsection{Effect of network area and energy}
Now by changing the parameters as shown in TABLE 2, Fig. 12 to Fig. 17 clearly show that, by increasing the area AM-DisCNT still perform far better than LEACH and DEEC.

 \begin{table}[htbp]
  \centering
  \caption{}
  \begin{tabular}[height=9cm,width=9cm]{lll}
    \toprule
    Parameter & Value  &  \\
    \midrule

   N            & 100  \\
  $E_0$          & 0.8 j \\
     $R1$        & 25m \\
     $R2$       & 40m \\

     \bottomrule
 \end{tabular}%
 \label{tab:addlabe2}%
 \end{table}%

\section{Conclusion and Future Work}
 AM-DisCNT is more energy efficient and effective in prolonging the network lifetime and stability of the network. AM-DisCNT is an energy aware protocol in wireless sensor networks. Unlike LEACH and DEEC, AM-DisCNT performs well in multi-level heterogeneous environment. Simulation results show that AM-DisCNT is more stable than LEACH and DEEC and offers greater lifetime and throughput.

 In future, we will implement different performance metrics of AM-DisCNT, as authors have done in ~\cite{17}, ~\cite{18}, ~\cite{19} and ~\cite{20}.

 \ifCLASSOPTIONcaptionsoff
  \newpage
\fi


\end{document}